\documentclass[aps,prl,pdf,superscriptaddress,amsmath,amssymb,amsfonts,twocolumn,showpacs,nofootinbib]{revtex4-2}

\usepackage{amssymb}
\usepackage{eqnarray,amsmath}
 
\usepackage{graphicx}
\usepackage{epsfig}
\usepackage{dcolumn}
\usepackage{bm}
\usepackage{braket}
\usepackage{amsmath}
\usepackage{balance}
\usepackage{physics}
\usepackage{csquotes}
\usepackage{mathtools}
\usepackage{graphicx,color,xcolor,colortbl}

\usepackage[colorlinks=true,
linkcolor=blue,
filecolor=blue,      
urlcolor=blue,
citecolor=blue]{hyperref}

\usepackage[normalem]{ulem}

\newcommand{\prlsection}[1]{%
  \noindent\textit{#1}---\ \ignorespaces}
\definecolor{Gray}{gray}{0.85}
\definecolor{LightCyan}{rgb}{0.88,1,1}

\newcolumntype{a}{>{\columncolor{Gray}}c}
\newcolumntype{b}{>{\columncolor{white}}c}

\begin{document}
\title{Vorticity-Crystalline Order Coupling in Supersolids: Excitations and Re-entrant Phases}
\author{M. Schubert}
\email{malte.schubert@fysik.lu.se}
\affiliation{Mathematical Physics and NanoLund, LTH, Lund University, Box 118, 22100 Lund, Sweden}

\author{K. Mukherjee}
\affiliation{Department of Engineering Science, University of Electro-Communications, Tokyo 182-8585, Japan}
\affiliation{Mathematical Physics and NanoLund, LTH,  Lund University, Box 118, 22100 Lund, Sweden}

\author{P. Stürmer}
\affiliation{Mathematical Physics and NanoLund, LTH, Lund University, Box 118, 22100 Lund, Sweden}

\author{S. M. Reimann}
\affiliation{Mathematical Physics and NanoLund, LTH, Lund University, Box 118, 22100 Lund, Sweden}
\begin{abstract}

Rotation is a natural tool in ultracold gases to break time-reversal symmetry, yet its impact on the collective excitations of supersolids remains largely unexplored. We show theoretically that tuning the rotation frequency, rather than the interparticle interactions, can trigger the superfluid-to-supersolid transition in Bose-Einstein condensates (dBECs). Computing excitation spectra in the presence of vortices and persistent currents, we uncover a vortex-driven \textit{de-softening} mechanism whereby quantized vorticity elevates the gapless Goldstone mode to a finite-energy roton, restoring superfluidity. This effect results in  re-entrant supersolid phases as a function of rotation frequency, revealing a fundamental coupling between topological defects and crystalline order.
\end{abstract}

\date{\today}
\maketitle

Breaking time-reversal symmetry (TRS) underlies diverse phenomena from cosmological matter-antimatter asymmetry \cite{Sakharov1967, Christenson1964} to chiral superconductivity \cite{Mackenzie2003}. In condensed matter, magnetic fields break TRS and lift Kramers degeneracy \cite{Kramers1930}, enabling the quantum Hall effect \cite{Klitzing1980, Laughlin1981} and access to nontrivial topological states \cite{Hasan2010, Qi2011}. Additionally, the competition between TRS-breaking fields and internal interactions often drives complex phase topographies, such as re-entrant superconductivity \cite{Meul1984} and fractional quantum Hall transitions \cite{Eisenstein2002}. Analogously, in ultracold Bose and Fermi gases, external rotation breaks TRS by acting as a synthetic magnetic field \cite{Fetter2009, Dalibard2011}. This induces the well-known hallmarks of superfluidity such as  quantized vortices~\cite{Chevy2000, *AboShaeer2001, *Engels2003, Zwierlein2005} and persistent currents~\cite{ Ramanathan2011, *Moulder2012, Del_pace2022,  *cai2022_Fermi}, offering a highly controllable platform for exploring TRS-broken effects~\cite{Mukherjee_BHS_2022, Yao2024}.

Externally applied rotating magnetic fields have made it possible to nucleate vortices in dBECs in the superfluid (SF) as well as the emergent supersolid (SS) phases \cite{Klaus2022, Casotti2024}. The SS phase \cite{Gross1957, Yang1962, Andreev1969, Chester1970, Boninsegni2012}, defined by the simultaneous breaking of global gauge and continuous translational symmetries, has now been realized in several ultracold platforms~\cite{Chomaz2023, mukherjee2023droplets, recati2023supersolidity} including lanthanide atomic gases \cite{Chomaz2019, Bottcher2019, Tanzi2019, Norcia2021}, light-mediated cavity systems \cite{Leonard2017}, and spin-orbit–coupled Bose gases \cite{Li2017}. Microscopically, the transition occurs due to the softening of a finite-momentum roton mode~\cite{Pomeau1994, Santos2003, Wilson2008, Blakie2012_Roton_a, Blakie2012_Roton_b, Chomaz2018, Schmidt2021}, which signals the shift from a homogeneous SF to a density-modulated SS state.

While such softening could be induced by flow velocities exceeding the Landau critical limit in superfluid $^{4}\mathrm{He}$ \cite{Landau1941_critical_velocity,  *Pitaevskii1984, *Ancilotto_He_2005},  current dipolar experiments~\cite{Ferrier-Barbut2016, Schmitt2016, Chomaz2016, Chomaz2018, Chomaz2019, Bottcher2019, Tanzi2019, Schmidt2021} typically access the roton instability through the precise tuning of anisotropic interparticle interactions instead of hydrodynamic flow. The resulting SS state comes with a rich hierarchy of collective excitations~\cite{Natale2019, Petter2019, Guo2019, Hertkorn2019, Hertkorn2021Spectrum2D, Ilg2023, kirkby2024, Poli2024Excitations2D}, including Higgs and Goldstone modes associated with spontaneous symmetry breaking \cite{Chomaz2018, Petter2019}. While these modes can reveal  key system properties such as the order of the phase transition~\cite{hertkorn2024decoupled}, they have so far been explored almost exclusively within the manifold of time-reversal-invariant states.

In this Letter, we theoretically uncover spectral signatures of TRS breaking in rotating dBECs, revealing how the interplay between rotation and long-range interactions reconfigures a superfluid phase. Combining analytical insight with numerical calculations of excitation spectra in geometries hosting vortices and persistent currents, we show that rotation can induce supersolidity in otherwise SF regimes. Crucially, vortex nucleation in the ensuing SS state can effectively \textit{de-soften} the Goldstone mode by lifting it back to a finite-energy roton excitation which restores the SF state. This novel mechanism yields a sequence of re-entrant supersolid pockets as a function of rotation frequency, where the discrete entry of vortices periodically suppresses and restores crystalline order. Our results open an experimentally feasible  route to probe the impact of TRS breaking on dipolar supersolids, revealing a hitherto unknown fundamental coupling between quantized vorticity and density modulation.

\indent \prlsection{Theoretical model} Let us consider a dilute, weakly interacting gas of ultracold dipolar atoms confined in a rotationally symmetric potential $V_{\rm trap}(\mathbf{r})$ and  subjected to an external rotation with angular velocity $\Omega$. The short-range contact and long-range dipolar potential are modelled by 
$
U_{\rm c} =
g\,\delta(\mathbf{r})$ with $g=\frac{4\pi\hbar^{2} a_{s}}{M}$
and 
$U_{\rm dd} = \frac{3\hbar^{2} a_{\mathrm{dd}}}{M}
\frac{1 - 3\cos^{2}\Theta}{r^{3}},
$ respectively. Here, $M$ is the atomic mass, $a_{s}$ is the tunable (via Feshbach resonance~\cite{chin2010,Tang2015a}) $s$-wave scattering 
length, and $a_{\mathrm{dd}}$ denotes the dipolar length. $\Theta$ is the angle 
between $\mathbf{r}$ and the polarization direction. To determine the ground states for a given $\Omega$, we solve the extended 
Gross–Pitaevskii equation (eGPE) for the condensate wavefunction 
$\Psi(\mathbf{r},t)$ in the rotating frame using imaginary-time propagation, including the 
Lee–Huang–Yang (LHY) beyond-mean-field correction in the 
local-density approximation~\cite{Lima2011, Lima2012, Wachtler2016}, characterized by the coefficient $g_{\rm QF}$ (additional information is provided in Appendix A). 

The excitation spectrum is obtained by linearizing the eGPE around the rotational ground state $\Psi_{0}$,
$\Psi(\mathbf{r}, t) = \Psi_{0}(\mathbf{r}) +  \delta\Psi(\mathbf{r}, t)$, using the Bogoliubov ansatz
$
\delta\Psi(\mathbf{r}, t) =
\left[
u_{j}(\mathbf{r})\, e^{-i\omega_{j} t}
+ v_{j}^{*}(\mathbf{r})\, e^{i\omega_{j} t}
\right] e^{-i\mu t/\hbar},
$
where $\mu$ is the chemical potential in the rotating frame. The linearization leads to the Bogoliubov--de Gennes (BdG) equations,
\begin{align}
\label{BdG}
    \begin{pmatrix}
    \hat{\mathcal{L}} & \hat{\mathcal{D}} \\
    -\hat{\mathcal{D}}^* & -\hat{\mathcal{L}}^*
    \end{pmatrix}
    \begin{pmatrix}
    u \\
    v
    \end{pmatrix}
    = \hbar\omega
    \begin{pmatrix}
    u \\
    v
    \end{pmatrix},
\end{align}
where $\hat{\mathcal{L}}u=\big[-\frac{\hbar^2}{2m}\nabla^2+V_{\rm trap}(\mathbf{r})
+2g|\Psi_0|^2+U_{\rm dd}*|\Psi_0|^2+\frac{5}{2}g_{\rm QF}|\Psi_0|^3
-\mu-\Omega\hat{L}_z\big]u
+\Psi_0 U_{\rm dd}*(\Psi_0^*u)$ and
$\hat{\mathcal{D}}u=\big[g\Psi_0^2+\frac{3}{2}g_{\rm QF}|\Psi_0|\Psi_0^2\big]u
+\Psi_0 U_{\rm dd}*(\Psi_0 u)$.
Solving these equations yields the excitation frequencies $\omega_j$,
the corresponding eigenfunctions $u_j$ and $v_j$, as well as the density
fluctuations of the ground state $
\delta n_j = 2\mathrm{Re}\!\left[(\Psi_{0}^*u_{j} + \Psi_0v_{j})e^{i \omega t}\right].
$

\begin{figure}
	\centering
	\includegraphics[width = 0.5\textwidth]{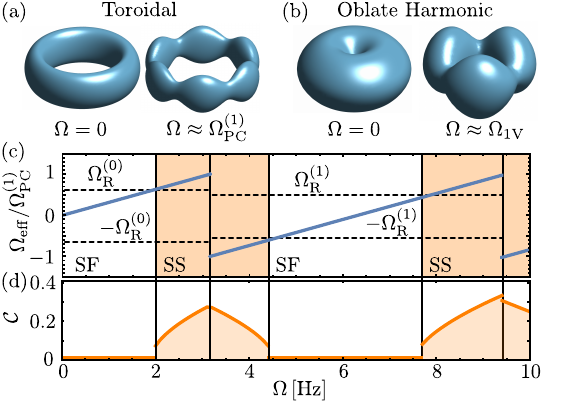}
    \vspace{-0.8cm}
   \caption{\textbf{Rotation-induced supersolidity.} (a)-(b) Three-dimensional density isosurfaces (at $40\%$ peak density) of the SF and SS states realized without and with external rotation $\Omega$, respectively, at fixed scattering length $a_s$, in (a) a toroidal and (b) an oblate harmonic trap. The scattering lengths are $a_s = 92.8a_0$ in (a) and $87.4a_0$ in (b), with critical rotation frequencies $\Omega_{\mathrm{PC}}^{(1)} \approx 3.2\,\mathrm{Hz}$ and $\Omega_{1\mathrm{V}} \approx 24.0\,\mathrm{Hz}$ marking the nucleation of a single-quantized persistent current and a unit vortex, respectively. (c) The effective hydrodynamic flow velocity $\Omega_{\mathrm{eff}} = \Omega-\Omega_{\mathrm{GS}}$ (solid blue line) versus external rotation $\Omega$ for a toroidal trap of radius $R$, where $\Omega_{\mathrm{GS}}=\hbar q/(MR^2)$ denotes the ground state angular velocity for a winding number $q$. The dashed black curve $\Omega_{\rm R}^{(q)}$ denotes the critical angular velocity for roton softening (or Goldstone de-softening). The vertical black lines signify $\Omega_{\mathrm{PC}}^{(q)}$, at which vortex nucleation reverses the effective flow, leading to a cyclic alternation between SF and SS phases as $\Omega$ varies. (d) Density contrast $\mathcal{C}$ of the ground state showing the re-entrant supersolid pockets.}
\label{figure1}
\end{figure}
In the following, we consider parameters relevant to bosonic
$^{162}\mathrm{Dy}$ atoms ($a_{\mathrm{dd}} = 130a_{0}$, with $a_{0}$ being the Bohr radius) and focus on angular excitation modes.
In our azimuthally symmetric geometry, where $[\hat{\mathcal{L}}, \hat{L}_z] = [\hat{\mathcal{D}}, \hat{L}_z] = 0$, the modes of a SF are characterized by well-defined angular-momentum quantum numbers $m$, corresponding to eigenstates of $\hat{L}_z$, and dominate the low-energy spectrum.  These excitations underpin the emergence of supersolidity, which we quantify via the density contrast $\mathcal{C} = (n_{\max}-n_{\min})/(n_{\max}+n_{\min})$, where $n_{\max}$ and $n_{\min}$ are the local maximum and minimum densities~\cite{Chomaz2023, Mukherjee_Selective2025}. In the uniform SF regime $\mathcal{C}\to0$, while $\mathcal{C}\to1$ indicates a fully modulated  state.

\indent \prlsection{Flow-induced asymmetry in the excitations}
To gain analytical insight into the angular excitations under rotation, we consider a toroidal geometry of radius $R$ under tight confinement ($R\gg\sqrt{\hbar/M\omega_r}$). In this limit, the critical angular velocity for nucleating a persistent current in a SF with winding number $q$ is
$\Omega_{\rm PC}^{(q)} = \hbar (q - \tfrac{1}{2})/(MR^{2})$, independent of the interactions.  Once such a current is established, the condensate acquires a finite circulation
characterized by the ground state angular velocity
$\Omega_{\rm GS} = \hbar q/(MR^{2})$. The resulting dynamics are governed by the effective hydrodynamic flow, defined by the relative angular velocity
$\Omega_{\mathrm{eff}} = \Omega - \Omega_{\rm GS}$. The corresponding excitation spectrum can be obtained by solving Eq.~\eqref{BdG} around the ground state $\Psi_0 \sim e^{i q \phi}$. Because the mode momentum $m$ couples to the persistent current, it is useful to introduce the quantum number $\mathcal{M}=m-q$, which is also the angular momentum of the density fluctuations $\delta n$. Then,  the excitation spectrum separates into two contributions and can be written as

\begin{align}
\label{Dispersion}
    \omega_{\mathcal{M}}^{(q)}(\Omega)= \tilde{\omega}^{(q)}_\mathcal{M} - \mathcal{M}\!\,\Omega_{\rm eff}.
\end{align}
The first term in Eq.~\eqref{Dispersion} incorporates interaction effects, including the long-range dipolar force and LHY corrections, and is independent of $\Omega$. As a result, it preserves momentum-reversal symmetry ($\mathcal{M} \rightarrow -\mathcal{M}$). Meanwhile, the second term introduces a Galilean shift that removes the energy degeneracy between counter-propagating modes. As a result, the spectral splitting is determined entirely by the effective flow: $\omega_{\mathcal{-M}}^{(q)}(\Omega)-\omega_{\mathcal{M}}^{(q)}(\Omega)=2\mathcal{M}\!\,\Omega_{\rm eff}$. This provides a direct spectral signature of the persistent current. For
$\Omega<\Omega_{\mathrm{PC}}^{(1)}$, the ground state carries no
circulation ($q=\Omega_{\mathrm{GS}}=0$), and the splitting follows the external rotation $\Omega$. When $\Omega>\Omega_{\mathrm{PC}}^{(1)}$, the system
transitions into the $q=1$ persistent-current state, causing the sign
of the effective flow $\Omega_{\mathrm{eff}}$ to reverse (see the thick blue line in Fig.~\ref{figure1}(c), which shows
$\Omega_{\mathrm{eff}}$ as a function of $\Omega$). This leads to mode swapping: the excitation mode that was previously lower in energy becomes the higher-energy one, and vice versa (see the subsequent discussion and Fig. \ref{Sup1} in Appendix B). As we will show, this exchange in energy ordering has a decisive impact on the crystalline order of the rotation-induced SS state.

\indent \prlsection{Roton instability and re-entrant SS state} Equation \eqref{Dispersion} provides a direct criterion for identifying the dynamical instabilities of the system. We focus on the regime where the long-range dipolar interaction induces a pronounced momentum dependence, driving a rotational instability via roton modes labeled by $m = m_{\rm R}$. In accordance with Eq.~\eqref{Dispersion}, the roton branches split: modes with $m_{\rm R}$ having the same sign as $\Omega$ are lowered in energy (roton$^{+}$), while those with the opposite sign are pushed to higher energies (roton$^{-}$). As $\Omega$ increases, the energy of the roton$^{+}$  softens until it vanishes at $\omega^{(0)}_{m_{\rm R}}(\Omega^{(0)}_{\rm R})=0$, defining the critical angular velocity $\Omega^{(0)}_{\rm R}$ for the roton instability in the absence of persistent current. For $\Omega>\Omega^{(0)}_{\rm R}$, the system enters a SS phase characterized by spontaneous density modulation. In this phase, the roton$^{+}$ mode becomes a gapless Goldstone mode, while the roton$^{-}$ mode evolves into a gapped Higgs$^{-}$ mode.

Entering the persistent current state at $\Omega > \Omega_{\mathrm{PC}}^{(1)}$ triggers the mode swapping. The previously counter-propagating Higgs$^{-}$ mode now  becomes a Higgs$^{+}$ mode that progressively softens with increasing $\Omega$. Once the effective velocity satisfies $ \Omega_{\mathrm{eff}} > -\Omega_{\rm R}^{(1)}$, this excitation converts into a roton$^{+}$ mode. Simultaneously, the zero-energy Goldstone mode evolves into a gapped roton$^{-}$ mode. The de-softening of the Goldstone mode into a finite-energy roton drives the transition from the modulated SS state back into a SF one. As $\Omega$ increases toward the $q=2$ threshold $\Omega_{\mathrm{PC}}^{(2)}$, the enhanced relative flow again triggers a roton softening and a recovery of the SS density modulation. Consequently, the sequential nucleation of persistent current states drives a cyclic phase evolution, which manifests as discrete pockets of supersolidity, see Fig.~\ref{figure1}(c)-(d), characterizing the re-entrant SS state.

\indent \prlsection{Numerical results for toroidal traps} We now support our analytical insights by numerically solving the eGPE and BdG equations, Eq.~\eqref{BdG}. First, we consider a toroidal trapping potential,
$V_{\mathrm{trap}}(\mathbf{r}) = M \omega_r^2 (\sqrt{x^2+y^2} - R)^2/2 + M \omega_z^2 z^2/2$,
with trap frequencies $(\omega_r, \omega_z)/2\pi = (100,120)\,\mathrm{Hz}$, radius $R = 2.8 \mu\mathrm{m}$, and atom number $N = 32400$. Dipolar BECs and their SS phases in toroidal geometries have so far only been studied theoretically~\cite{Abad2012a, Abad2012b, Tengstrand2021, Tengstrand2023, sindik2024, hertkorn2024decoupled, HiggsMukherjee2025, Schubert2025,Preti2025}; however, experimental realization is expected to follow soon. In the absence of rotation, our system undergoes a SF-SS transition at $a_s \approx 92.73a_0$, defining the quantum-critical point. Slightly above this threshold, we show the excitation spectra as a function of $\Omega$ at $a_s = 92.8a_0$ in Fig.~\ref{figure2}(a). 

At $\Omega = 0$, the spectrum exhibits a set of roton modes, along with higher-lying superfluid sound modes; see, for example, the profile ($f = u + v^{*}$) of the $m_{\rm R} = 6$ roton mode in Fig.~\ref{figure2}(a). Finite rotation lifts the degeneracy, causing the $m_{\rm R} = 6$ roton$^{+}$ (roton$^{-}$) branch to soften (harden). At $\Omega_{\rm R}^{(0)} \approx 2\pi \times 1.95$ Hz, the energy of the roton$^{+}$ mode vanishes, signaling the spontaneous breaking of azimuthal symmetry and the formation of a SS with $n_d = 6$ localized density sites; see also Fig.~\ref{figure1}(a). Here, the roton$^{+}$ mode evolves into a gapless Goldstone mode, while the roton$^{-}$ becomes a gapped Higgs$^{-}$ mode.

A key highlight of our work is the nature of the phase transition. As illustrated in Fig.~\ref{figure2}(b), the presence of a finite Higgs-mode gap at the critical point suggests a first-order character, which we explain within a Ginzburg-Landau framework~\cite{Zhang2019, Biagioni2022_order}. We express the energy in the rotating frame as $E' = E - L_z \Omega$. Taking the superfluid fraction $f_s$ as the order parameter and treating the angular momentum classically, we approximate $L_z \approx I \Omega (1 - f_s)$, where $I$ is the moment of inertia. This substitution introduces a term linear in $f_s$ into the energy functional. Consequently, minimizing $E'(f_s)$ leads to a discontinuous jump in $f_s$ at the transition. Crucially, while interaction-driven supersolid transitions in rings are typically second-order~\cite{hertkorn2024decoupled}, the rotational breaking of TRS fundamentally shifts the transition to first-order.

The red line in Fig.~\ref{figure2} shows the energy difference $E_{\rm PC}=E^{\prime}(q=1) - E^{\prime}(q=0)$, obtained from the eGPE, indicating that for $\Omega > \Omega_{\mathrm{PC}}^{(1)} \approx 2\pi \times 3.2$~Hz the state with $q=1$ becomes energetically favorable. Correspondingly, we observe the swapping of angular momenta of modes predicted by Eq.~\eqref{Dispersion} and the subsequent reduction of the energy of the Higgs$^{+}$ mode. At $\Omega_{\rm R}^{(1)} \approx 2\pi \times 4.47$~Hz, the system re-enters the SF phase, evidenced by the de-softening of the Goldstone mode. 
\begin{figure}[t]
	\centering
	\includegraphics[width = 0.486\textwidth]{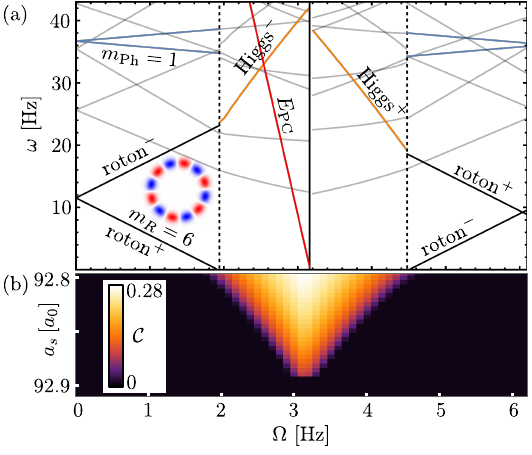}
    \vspace{-0.8cm}
\caption{\textbf{Excitation spectrum and phase diagram in toroidal geometry.} 
(a) BdG excitation energies vs.\ rotational velocity $\Omega$ at $a_s = 92.8a_{0}$. Black, blue, and orange lines denote the lowest roton ($m_{\rm R} = 6$), superfluid phonon ($m_{\rm Ph} = 1$), and Higgs modes, respectively. Vertical lines indicate the sequence of SF $\to$ SS, vortex nucleation, and SS $\to$ SF transitions. Signs $\pm$ indicate angular momentum orientation relative to $\Omega$. The red curve shows the persistent-current energy $E_{\rm PC}$. 
(Inset) Bogoliubov mode profile $f$ for the $m_{\rm R} = 6$ roton at $\Omega = 0$ in the $xy$-plane (red/blue indicates negative/positive amplitude). 
(b) Density contrast $\mathcal{C}$ in the $(\Omega, a_s)$ plane, delineating the SF and SS regimes. See text for other parameters.}
\label{figure2}

\label{figure2}
\end{figure}
As $\Omega$ increases, this cycle repeats, generating a sequence of SF $\to$ SS $\to$ SF transitions [Fig.~\ref{figure1}(d)]. This periodic recurrence, a re-entrant supersolidity,  constitutes the main result of this work. While re-entrance in ultracold gases has been mostly studied in optical lattices~\cite{KleinertRenetrant2004, Shenreentarnt2012, resupersolid_lattice2023, Piekarska2023, Krzywicka2024, Ding2025}, the phenomenon reported here is fundamentally distinct; it is driven by the discrete nucleation of topological charge, which periodically restores and suppresses crystalline order. Since the critical velocities for roton softening and Goldstone de-softening depend on the winding number $q$, the rotation-induced SS phase is restricted to the intervals $|\Omega - \Omega_{\mathrm{GS}}| < \Omega_{\rm R}^{(q)}$.
 Notably, a phononic instability can also occur in the $m = m_{\mathrm{ph}} = 1$ mode at a critical angular velocity $\Omega_{\mathrm{Ph}} = \tilde{\omega}_1$. However, for a rotation-induced supersolid to emerge, the roton instability must occur first, requiring $\Omega_{\rm R}^{(q)} < \Omega_{\mathrm{Ph}}^{(q)}$. This condition is satisfied near the quantum-critical point, where the roton energies lie below phononic ones (see the blue curve in Fig.~\ref{figure2}(a)).

\begin{figure}[t]
	\centering
	\includegraphics[width = 0.49\textwidth]{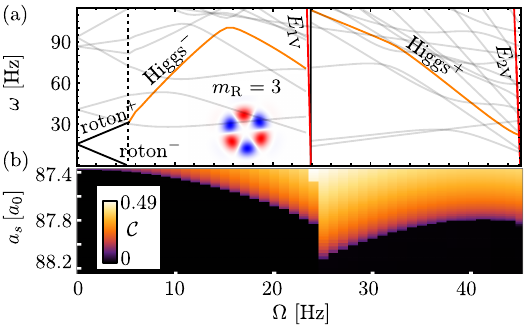}
    \vspace{-0.7cm}
\caption{\textbf{Excitation spectrum and phase diagram  in an
oblate harmonic trap.}
(a) Excitation frequencies of rotons (black), Higgs (orange), and all other modes (gray) versus rotational angular velocity $\Omega$. The dashed and solid vertical lines indicate $\Omega_{\rm R}^{(0)}$ and $\Omega_{1\rm V}$, respectively, while the red lines denote the energy differences of the one- and two-vortex states relative to the ground state, $E_{1\rm V}$ and $E_{2\rm V}$.  Inset: mode profile $f$ of the $m_{\rm R}=3$ roton driving the SF$\rightarrow$SS transition.
(b) Density-modulation contrast $\mathcal{C}$ as a function of $\Omega$ and the $s$-wave scattering length $a_s$. Other parameters are given in the text.}

\label{figure3}
\end{figure}
 In Fig.~\ref{figure2}(b), we map the existence of the rotation-induced supersolid phase in the $(\Omega, a_s)$ parameter space using the density contrast $\mathcal{C}$. A finite $\mathcal{C}$, signaling SS formation, emerges over a broader range of $\Omega$ near the quantum critical point and a narrower range away from it. The maximum of $\mathcal{C}$ occurs at $\Omega = \Omega_{\mathrm{PC}}^{(1)} = 2\pi \times 3.2\,\mathrm{Hz}$, in agreement with our analytical predictions. Since $\Omega_{\mathrm{PC}}^{(1)}$ increases with decreasing ring radius $R$, this provides a route to stabilize the SS phase over a wider range of $\Omega$.

\begin{figure}[b]
	\centering
	\includegraphics[width = 0.48\textwidth]{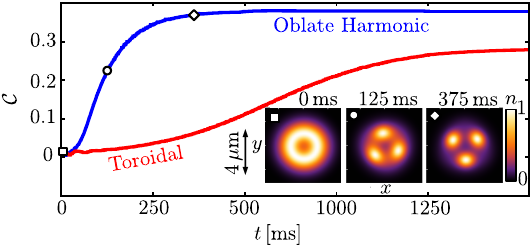}
    \vspace{-0.6 cm}
\caption{\textbf{Dynamical onset of supersolid order.} Time evolution of the density contrast $\mathcal{C}$ for toroidal (red) and oblate harmonic (blue) traps. Following a sudden quench of rotation to $\Omega/2\pi = 3$~Hz (toroidal) and $23$~Hz (oblate harmonic), the nonrotating SF ground state undergoes spontaneous density modulation. (Insets) Snapshots of the integrated density $n(x,y)$ in the oblate harmonic trap at specific time intervals (see markers), refer to Ref. \cite{Suplementary} for video of the dynamics. Scattering lengths are $a_s = 92.8\,a_0$ (toroidal) and $87.4\,a_0$ (oblate harmonic).}
\label{figure4}
\end{figure}

\prlsection{Numerical results for oblate harmonic traps}
To demonstrate that the rotationally induced SS phase is not restricted to specific geometries and to align our findings with established platforms~\cite{Schmidt2021OblateRoton, Casotti2024,Molignini2025}, we next consider an oblate harmonic trap. We set $R=0$ with trapping frequencies $(\omega_r, \omega_z) = 2\pi \times (125, 250)$~Hz for a system containing $N = 2 \times 10^{4}$ atoms~\cite{Hertkorn2021Spectrum2D}. The resulting density isosurfaces for both non-rotating SF and rotating SS ground states are displayed in Fig.~\ref{figure1}(b) above. 

At $\Omega = 0$, the SF exhibits a density minimum at the trap center while preserving azimuthal symmetry. This configuration supports angular roton modes as the lowest-energy excitations, analogous to the toroidal geometry. Under rotation, the degeneracy of these modes is lifted, causing the branches to split and soften. Specifically, the $m_{\rm R} = 3$ roton drives the transition to a SS state characterized by three distinct density maxima. This transition is marked by the emergence of zero-energy Goldstone and Higgs modes, as illustrated in Fig.~\ref{figure3}(a). Note that we only highlight the roton and Higgs modes in Fig. \ref{figure3}(a), while plotting the rest of the spectrum in gray.

Notably, the finite central density in the harmonic trap renders vortex nucleation energetically more expensive than in the toroidal case~\cite{Roccuzzo2020, Marija2022Vortex, Casotti2024}. This shifts the critical rotation frequencies for vortex formation to higher values [see the red lines in Fig.~\ref{figure3}(a)]. Consequently, the parameter space supporting the rotation-induced SS phase is significantly expanded, as shown in the phase diagram in Fig.~\ref{figure3}(b). For scattering lengths $a_s \lesssim 87.8 \, a_0$, we find that $\Omega_{\rm R}^{(1)} < \Omega_{1\rm V}$, indicating that the roton instability occurs before the onset of vortex entry. In contrast, for $a_s \gtrsim 87.8 \, a_0$, this ordering is reversed ($\Omega_{\rm R}^{(1)} > \Omega_{1\rm V}$). In the narrow intermediate regime $87.8 \, a_0 \lesssim a_s \lesssim 88.1 \, a_0$, the roton mode exhibits a characteristic de-softening behavior. Although mode coupling makes the rotational excitation spectra more intricate than in the toroidal case, the essential features of rotation-induced softening of the roton and de-softening of the Goldstone mode remain robust, as evident in Fig. \ref{figure3}.

\prlsection{Dynamical emergence of SS under rotation}
To establish the experimental feasibility of our results, we perform real-time simulations incorporating phenomenological damping $\gamma$ and  noise in the initial state. Starting from a non-rotating SF ground state, we suddenly quench the rotation at $t=0$ to $\Omega/2\pi = 23$~Hz for the oblate harmonic trap. This value sits slightly below the critical frequency $\Omega^{(1)}_{\mathrm{PC}}/2\pi = 24$~Hz and precludes vortex nucleation while inducing pronounced azimuthal symmetry breaking. Figure~\ref{figure4} illustrates the time evolution of the density contrast $\mathcal{C}$ (blue line), which grows from zero and saturates at $\mathcal{C} \approx 0.38$, signaling the formation of a stable SS phase. The insets in Fig.~\ref{figure4} display snapshots of the integrated 2D density profiles, $n(x,y) = \int |\Psi|^2\,dz$, at selected time intervals (indicated by markers), capturing the emergence of the crystalline structure. An analogous dynamical evolution occurs for the toroidal geometry (red line in Fig.~\ref{figure4}), demonstrating the robustness of the rotationally induced SS across different trapping potentials.

\prlsection{Conclusions and Outlook}
We have investigated the Bogoliubov excitation spectra of rotating dipolar condensates. In geometries supporting angular excitations, such as toroidal and oblate harmonic traps, the breaking of TRS lifts roton degeneracy and triggers a first-order SF-SS transition. We uncover a nontrivial interplay between crystalline order and topological excitations, namely, persistent currents and quantized vortices, which drives a periodic sequence of SF $\leftrightarrow$ SS transitions. This behavior is mediated by the de-softening of the gapless Goldstone mode and the simultaneous suppression of density modulations, giving rise to re-entrant SS phases, a phenomenon widely encountered in diverse systems such as liquid crystals~\cite{Cladis1975}, spin glasses~\cite{Gabay1981, *Aeppli1982}, quantum hall states~\cite{Cooper1999} and superconductivity~\cite{Jaccarino1962}.

Our work not only reveals an alternative to interaction-driven mechanisms for supersolidity via the excitation spectrum, but also paves the way for future studies of re-entrant phases in dBECs. Notably, these could be  observed in diverse platforms, including dipolar molecules~\cite{zhang2025observation, *zhang2025supersolid} and exciton–polariton condensates~\cite{Manni2011, *Trypogeorgos2025}, and other lanthanide elements~\cite{Chomaz2019, Miyazawa2022}, provided that: (i) the Hamiltonian supports low-energy angular excitations; (ii) the spectrum is concave; and (iii) the roton softening threshold $\Omega_{\rm R}$ is the system's lowest critical velocity. Utilizing rotation as a control parameter, our work enables SS in regimes where relatively weak long-range interactions would otherwise suppress it. Immediate extensions of this work include exploring finite-temperature effects~\cite{Sanchez-Baena2023} and further illuminating the other excitations, such as sound modes~\cite{Tomasz2025}, revealing novel Josephson oscillations~\cite{biagioni2024measurement,Platt2024} and shear waves~\cite{SenarathShear2025} within the SS.

{\it Acknowledgments}---We are grateful to Tilman Pfau and Ralf Klemt, and the Stuttgart dysprosium team, as well as Tiziano Arnone Cardinale, Deepak Gaur and Lila Chergui for insightful discussions. Financial support from the Knut and Alice Wallenberg Foundation (Grant No.~KAW2023.0322) and the Swedish Research Council (Grant No.~2022-03654 VR) is acknowledged. K.~M. acknowledges financial support through the JSPS Postdoctoral Fellowship (Fellowship No.~P25029).

\bibliography{reference}

@article{Kramers1930,
  author  = {Kramers, H. A.},
  title   = {La th{\'e}orie du pouvoir rotatoire paramagn{\'e}tique},
  journal = {Proc. Acad. Sci. Amsterdam},
  volume  = {33},
  pages   = {959},
  year    = {1930},
}

@article{Klitzing1980,
  author  = {von Klitzing, K. and Dorda, G. and Pepper, M.},
  title   = {New Method for High-Precision Determination of the Fine-Structure Constant Based on Quantized Hall Resistance},
  journal = {Phys. Rev. Lett.},
  volume  = {45},
  pages   = {494--497},
  year    = {1980},
  doi     = {10.1103/PhysRevLett.45.494},
  url     = {https://journals.aps.org/prl/abstract/10.1103/PhysRevLett.45.494}
}

@article{Laughlin1981,
  author  = {Laughlin, R. B.},
  title   = {Quantized {Hall} conductivity in two dimensions},
  journal = {Phys. Rev. B},
  volume  = {23},
  pages   = {5632--5633},
  year    = {1981},
  doi     = {10.1103/PhysRevB.23.5632},
  url     = {https://journals.aps.org/prb/abstract/10.1103/PhysRevB.23.5632}
}

@article{Christenson1964,
  author  = {Christenson, J. H. and Cronin, J. W. and Fitch, V. L. and Turlay, R.},
  title   = {Evidence for the $2\pi$ Decay of the $K_2^0$ Meson},
  journal = {Phys. Rev. Lett.},
  volume  = {13},
  pages   = {138--140},
  year    = {1964},
  doi     = {10.1103/PhysRevLett.13.138},
  url     = {https://journals.aps.org/prl/abstract/10.1103/PhysRevLett.13.138}
}

@article{Sakharov1967,
  author  = {Sakharov, A. D.},
  title   = {Violation of CP invariance, C asymmetry, and baryon asymmetry of the universe},
  journal = {Pisma Zh. Eksp. Teor. Fiz.},
  volume  = {5},
  pages   = {32},
  year    = {1967},
  note    = {[JETP Lett. 5, 24 (1967)]},
  url     = {https://inspirehep.net/files/61b1b5f39b3f1a22e6c5a8a7c3c7a9d6}
}

@article{Mackenzie2003,
  author  = {Mackenzie, Andrew Peter and Maeno, Yoshiteru},
  title   = {The superconductivity of $\text{Sr}_2\text{Ru}\text{O}_4$ and the physics of spin-triplet pairing},
  journal = {Rev. Mod. Phys.},
  volume  = {75},
  pages   = {657--712},
  year    = {2003},
  doi     = {10.1103/RevModPhys.75.657},
  url     = {https://journals.aps.org/rmp/abstract/10.1103/RevModPhys.75.657}
}

@article{Hasan2010,
  author  = {Hasan, M. Z. and Kane, C. L.},
  title   = {Colloquium: Topological insulators},
  journal = {Rev. Mod. Phys.},
  volume  = {82},
  pages   = {3045--3067},
  year    = {2010},
  doi     = {10.1103/RevModPhys.82.3045},
  url     = {https://journals.aps.org/rmp/abstract/10.1103/RevModPhys.82.3045}
}

@article{Qi2011,
  author  = {Qi, Xiao-Liang and Zhang, Shou-Cheng},
  title   = {Topological insulators and superconductors},
  journal = {Rev. Mod. Phys.},
  volume  = {83},
  pages   = {1057--1110},
  year    = {2011},
  doi     = {10.1103/RevModPhys.83.1057},
  url     = {https://journals.aps.org/rmp/abstract/10.1103/RevModPhys.83.1057}
}

@article{Marija2022Vortex,
  title = {Creation and robustness of quantized vortices in a dipolar supersolid when crossing the superfluid-to-supersolid transition},
  author = {\ifmmode \check{S}\else \v{S}\fi{}indik, Marija and Recati, Alessio and Roccuzzo, Santo Maria and Santos, Luis and Stringari, Sandro},
  journal = {Phys. Rev. A},
  volume = {106},
  issue = {6},
  pages = {L061303},
  numpages = {5},
  year = {2022},
  month = {Dec},
  publisher = {American Physical Society},
  doi = {10.1103/PhysRevA.106.L061303},
  url = {https://link.aps.org/doi/10.1103/PhysRevA.106.L061303}
}

@article{Jaccarino1962,
  title = {Ultra-High-Field Superconductivity},
  author = {Jaccarino, V. and Peter, M.},
  journal = {Phys. Rev. Lett.},
  volume = {9},
  issue = {7},
  pages = {290--292},
  year = {1962},
  publisher = {American Physical Society},
  doi = {10.1103/PhysRevLett.9.290}
}

@article{Meul1984,
  title = {Observation of Magnetic-Field-Induced Superconductivity},
  author = {Meul, H. W. and Rossel, C. and Decroux, M. and Fischer, \O{}. and Remenyi, G. and Briggs, A.},
  journal = {Phys. Rev. Lett.},
  volume = {53},
  issue = {5},
  pages = {497--500},
  year = {1984},
  publisher = {American Physical Society},
  doi = {10.1103/PhysRevLett.53.497}
}

@article{Cladis1975,
  title = {New Liquid-Crystal Phase Diagram: An Isotropic-Nematic-Smectic-$A$-Nematic Transitions},
  author = {Cladis, P. E.},
  journal = {Phys. Rev. Lett.},
  volume = {35},
  issue = {1},
  pages = {48--51},
  year = {1975},
  publisher = {American Physical Society},
  doi = {10.1103/PhysRevLett.35.48}
}

@article{Eisenstein2002,
  title = {Insulating and Fractional Quantum {Hall} States in the First Excited {Landau} Level},
  author = {Eisenstein, J. P. and Lewis, K. B. and Yeh, A. S. and West, K. W. and Pfeiffer, L. N.},
  journal = {Phys. Rev. Lett.},
  volume = {88},
  issue = {7},
  pages = {076801},
  year = {2002},
  publisher = {American Physical Society},
  doi = {10.1103/PhysRevLett.88.076801}
}

@article{Cooper1999,
  title = {Insulating phases of two-dimensional electrons in high Landau levels: Observation of sharp thresholds to conduction},
  author = {Cooper, K. B. and Lilly, M. P. and Eisenstein, J. P. and Pfeiffer, L. N. and West, K. W.},
  journal = {Phys. Rev. B},
  volume = {60},
  issue = {16},
  pages = {R11285--R11288},
  year = {1999},
  publisher = {American Physical Society},
  doi = {10.1103/PhysRevB.60.R11285}
}

@article{Gabay1981,
  title = {Coexistence of Spin-Glass and Ferromagnetic Orderings},
  author = {Gabay, M. and Toulouse, G.},
  journal = {Phys. Rev. Lett.},
  volume = {47},
  issue = {3},
  pages = {201--204},
  year = {1981},
  publisher = {American Physical Society},
  doi = {10.1103/PhysRevLett.47.201}
}

@article{Aeppli1982,
  title = {Spin-wave dynamics in a reentrant spin-glass},
  author = {Aeppli, G. and Shapiro, S. M. and Birgeneau, R. J. and Chen, H. S.},
  journal = {Phys. Rev. B},
  volume = {25},
  issue = {7},
  pages = {4882--4885},
  year = {1982},
  publisher = {American Physical Society},
  doi = {10.1103/PhysRevB.25.4882}
}

@article{Tomasz2025,
  title = {Anomalous Doppler Effect in Superfluid and Supersolid Atomic Gases},
  author = {Zawi\ifmmode \acute{s}\else \'{s}\fi{}lak, Tomasz and \ifmmode \check{S}\else \v{S}\fi{}indik, Marija and Stringari, Sandro and Recati, Alessio},
  journal = {Phys. Rev. Lett.},
  volume = {134},
  issue = {22},
  pages = {226001},
  numpages = {7},
  year = {2025},
  month = {Jun},
  publisher = {American Physical Society},
  doi = {10.1103/PhysRevLett.134.226001},
  url = {https://link.aps.org/doi/10.1103/PhysRevLett.134.226001}
}

@article{Miyazawa2022,
  title = {Bose-Einstein Condensation of Europium},
  author = {Miyazawa, Yuki and Inoue, Ryotaro and Matsui, Hiroki and Nomura, Gyohei and Kozuma, Mikio},
  journal = {Phys. Rev. Lett.},
  volume = {129},
  issue = {22},
  pages = {223401},
  numpages = {5},
  year = {2022},
  month = {Nov},
  publisher = {American Physical Society},
  doi = {10.1103/PhysRevLett.129.223401},
  url = {https://link.aps.org/doi/10.1103/PhysRevLett.129.223401}
}

@article{Mukherjee_Selective2025,
  title = {Selective rotation and attractive persistent currents in antidipolar ring supersolids},
  author = {Mukherjee, K. and Cardinale, T. Arnone and Reimann, S. M.},
  journal = {Phys. Rev. A},
  volume = {111},
  issue = {3},
  pages = {033304},
  numpages = {9},
  year = {2025},
  month = {Mar},
  publisher = {American Physical Society},
  doi = {10.1103/PhysRevA.111.033304},
  url = {https://link.aps.org/doi/10.1103/PhysRevA.111.033304}
}

@article{Biagioni2022_order,
  title = {Dimensional Crossover in the Superfluid-Supersolid Quantum Phase Transition},
  author = {Biagioni, Giulio and Antolini, Nicol\`o and Ala\~na, Aitor and Modugno, Michele and Fioretti, Andrea and Gabbanini, Carlo and Tanzi, Luca and Modugno, Giovanni},
  journal = {Phys. Rev. X},
  volume = {12},
  issue = {2},
  pages = {021019},
  numpages = {18},
  year = {2022},
  month = {Apr},
  publisher = {American Physical Society},
  doi = {10.1103/PhysRevX.12.021019},
  url = {https://link.aps.org/doi/10.1103/PhysRevX.12.021019}
}

@article{Landau1941_critical_velocity,
  title = {Theory of the Superfluidity of Helium II},
  author = {Landau, L.},
  journal = {Phys. Rev.},
  volume = {60},
  issue = {4},
  pages = {356--358},
  numpages = {0},
  year = {1941},
  month = {Aug},
  publisher = {American Physical Society},
  doi = {10.1103/PhysRev.60.356},
  url = {https://link.aps.org/doi/10.1103/PhysRev.60.356}
}

@article{Ancilotto_He_2005,
  title = {Density pattern in supercritical flow of liquid $^{4}\mathrm{He}$},
  author = {Ancilotto, F. and Dalfovo, F. and Pitaevskii, L. P. and Toigo, F.},
  journal = {Phys. Rev. B},
  volume = {71},
  issue = {10},
  pages = {104530},
  numpages = {5},
  year = {2005},
  month = {Mar},
  publisher = {American Physical Society},
  doi = {10.1103/PhysRevB.71.104530}
}

@article{SenarathShear2025,
  title = {Anomalous dispersion of shear waves in dipolar supersolids},
  author = {Senarath Yapa, P. and Bland, T.},
  journal = {Phys. Rev. A},
  volume = {112},
  issue = {2},
  pages = {L021303},
  numpages = {8},
  year = {2025},
  month = {Aug},
  publisher = {American Physical Society},
  doi = {10.1103/xz57-52ft},
  url = {https://link.aps.org/doi/10.1103/xz57-52ft}
}

@article{hertkorn2024decoupled,
  title={{Decoupled sound and amplitude modes in trapped dipolar supersolids}},
  author={Hertkorn, Jens and St{\"u}rmer, Philipp and Mukherjee, Koushik and Ng, Kevin SH and Uerlings, Paul and Hellstern, Fiona and Lavoine, Lucas and Reimann, SM and Pfau, Tilman and Klemt, Ralf},
  journal={Physical Review Research},
  volume={6},
  number={4},
  pages={L042056},
  year={2024},
  publisher={APS},
url = {https://doi.org/10.1103/PhysRevResearch.6.L042056}
}

@article{chin2010,
  title = {Feshbach resonances in ultracold gases},
  author = {Chin, Cheng and Grimm, Rudolf and Julienne, Paul and Tiesinga, Eite},
  journal = {Rev. Mod. Phys.},
  volume = {82},
  issue = {2},
  pages = {1225--1286},
  numpages = {0},
  year = {2010},
  month = {Apr},
  publisher = {American Physical Society},
  doi = {10.1103/RevModPhys.82.1225},
  url = {https://link.aps.org/doi/10.1103/RevModPhys.82.1225}
}

@article{recati2023supersolidity,
  title={Supersolidity in ultracold dipolar gases},
  author={Recati, Alessio and Stringari, Sandro},
  journal={Nature Reviews Physics},
  volume={5},
  number={12},
  pages={735--743},
  year={2023},
  publisher={Nature Publishing Group UK London},
  url = {https://doi.org/10.1038/s42254-023-00648-2}
}

@article{Molignini2025,
doi = {10.1088/1361-648X/ae0fd3},
url = {https://doi.org/10.1088/1361-648X/ae0fd3},
year = {2025},
month = {nov},
publisher = {IOP Publishing},
volume = {37},
number = {44},
pages = {445401},
author = {Molignini, Paolo},
title = {Beyond-mean-field phases of rotating dipolar condensates in the strongly correlated regime},
journal = {Journal of Physics: Condensed Matter}
}

@article{Schubert2025,
  title = {Josephson vortices and persistent current in a double-ring supersolid system},
  author = {Schubert, M. and Mukherjee, K. and Pfau, T. and Reimann, S. M.},
  journal = {Phys. Rev. Res.},
  volume = {7},
  issue = {3},
  pages = {033110},
  numpages = {13},
  year = {2025},
  month = {Aug},
  publisher = {American Physical Society},
  doi = {10.1103/tl7c-v5bs},
  url = {https://link.aps.org/doi/10.1103/tl7c-v5bs}
}

@article{Fetter2009,
  title = {Rotating trapped {Bose-Einstein} condensates},
  author = {Fetter, Alexander L.},
  journal = {Rev. Mod. Phys.},
  volume = {81},
  issue = {2},
  pages = {647--691},
  year = {2009},
  doi = {10.1103/RevModPhys.81.647}
}

@article{Dalibard2011,
  title = {Colloquium: Artificial gauge potentials for neutral atoms},
  author = {Dalibard, Jean and Gerbier, Fabrice and Juzeli\'{n}as, Gediminas and {\"O}hberg, Patrik},
  journal = {Rev. Mod. Phys.},
  volume = {83},
  issue = {4},
  pages = {1523--1543},
  year = {2011},
  doi = {10.1103/RevModPhys.83.1523}
}

@article{Chevy2000,
  title = {Measurement of the Angular Momentum of a Rotating Bose-Einstein Condensate},
  author = {Chevy, F. and Madison, K. W. and Dalibard, J.},
  journal = {Phys. Rev. Lett.},
  volume = {85},
  issue = {11},
  pages = {2223--2227},
  numpages = {0},
  year = {2000},
  month = {Sep},
  publisher = {American Physical Society},
  doi = {10.1103/PhysRevLett.85.2223},
  url = {https://link.aps.org/doi/10.1103/PhysRevLett.85.2223}
}

@article{AboShaeer2001,
  author  = {Abo-Shaeer, J. R. and Raman, C. and Vogels, J. M. and Ketterle, W.},
  title   = {Observation of Vortex Lattices in {Bose-Einstein Condensates}},
  journal = {Science},
  volume  = {292},
  number  = {5516},
  pages   = {476--479},
  year    = {2001},
  doi     = {10.1126/science.1060182},
  url     = {https://www.science.org/doi/10.1126/science.1060182}
}

@article{Engels2003,
  title = {Observation of Long-Lived Vortex Aggregates in Rapidly Rotating {Bose-Einstein Condensates}},
  author = {Engels, P. and Coddington, I. and Haljan, P. C. and Schweikhard, V. and Cornell, E. A.},
  journal = {Phys. Rev. Lett.},
  volume = {90},
  issue = {17},
  pages = {170405},
  numpages = {4},
  year = {2003},
  month = {May},
  publisher = {American Physical Society},
  doi = {10.1103/PhysRevLett.90.170405},
  url = {https://link.aps.org/doi/10.1103/PhysRevLett.90.170405}
}

@article{Zwierlein2005,
  author  = {Zwierlein, M. W. and Abo-Shaeer, J. R. and Schunck, C. H. and Schirotzek, A. and Ketterle, W.},
  title   = {Vortices and Superfluidity in a Strongly Interacting {Fermi} Gas},
  journal = {Nature},
  volume  = {435},
  pages   = {1047--1051},
  year    = {2005},
  doi     = {10.1038/nature03858},
  url     = {https://www.nature.com/articles/nature03858}
}

@article{Ramanathan2011,
  title = {Superflow in a Toroidal Bose-Einstein Condensate: An Atomic SQUID},
  author = {Ramanathan, A. and Wright, K. C. and Muniz, S. R. and Zelan, M. and Hill, W. T. and Lobb, C. J. and Helmerson, K. and Phillips, W. D. and Campbell, G. K.},
  journal = {Phys. Rev. Lett.},
  volume = {106},
  pages = {130401},
  year = {2011},
  doi = {10.1103/PhysRevLett.106.130401}
}

@article{Moulder2012,
  title = {Quantized supercurrents in annular {Bose-Einstein} condensates},
  author = {Moulder, S. and Beattie, S. and Smith, R. P. and Tammuz, N. and Hadzibabic, Z.},
  journal = {Phys. Rev. A},
  volume = {86},
  pages = {013603},
  year = {2012},
  doi = {10.1103/PhysRevA.86.013603}
}

@article{KleinertRenetrant2004,
  title = {Reentrant Phenomenon in the Quantum Phase Transitions of a Gas of Bosons Trapped in an Optical Lattice},
  author = {Kleinert, H. and Schmidt, S. and Pelster, A.},
  journal = {Phys. Rev. Lett.},
  volume = {93},
  issue = {16},
  pages = {160402},
  numpages = {4},
  year = {2004},
  month = {Oct},
  publisher = {American Physical Society},
  doi = {10.1103/PhysRevLett.93.160402},
  url = {https://link.aps.org/doi/10.1103/PhysRevLett.93.160402}
}

@article{Shenreentarnt2012,
  title = {Reentrant BCS-BEC Crossover and a Superfluid-Insulator Transition in Optical Lattices},
  author = {Shen, Zhaochuan and Radzihovsky, L. and Gurarie, V.},
  journal = {Phys. Rev. Lett.},
  volume = {109},
  issue = {24},
  pages = {245302},
  numpages = {5},
  year = {2012},
  month = {Dec},
  publisher = {American Physical Society},
  doi = {10.1103/PhysRevLett.109.245302},
  url = {https://link.aps.org/doi/10.1103/PhysRevLett.109.245302}
}

@article{Krzywicka2024,
  author  = {Krzywicka, A. and Polak, T. P.},
  title   = {Reentrant phase behavior in systems with density-induced tunneling},
  journal = {Sci. Rep.},
  volume  = {14},
  pages   = {10364},
  year    = {2024},
  doi     = {10.1038/s41598-024-61136-8},
  url     = {https://www.nature.com/articles/s41598-024-61136-8}
}

@article{Piekarska2023,
  title   = {Reentrant phase transitions involving glassy and superfluid orders in the random hopping Bose--Hubbard model},
  author  = {Piekarska, Anna M. and Kope\'c, Tadeusz K.},
  journal = {Physica A: Statistical Mechanics and its Applications},
  volume  = {609},
  pages   = {128360},
  year    = {2023},
  doi     = {10.1016/j.physa.2022.128360},
  url     = {https://www.sciencedirect.com/science/article/pii/S0378437122009189}
}

@misc{Suplementary,
  title={See supplementary material for a video of the full dynamics, which show the density isosurfaces in the rotating frame for the oblate harmonic trap taken at $20\%$ and $40\%$ of the maximum density}
}

@article{resupersolid_lattice2023,
  title={Reentrant supersolidity},
  author={Leo Radzihovsky},
  journal={	arXiv:2311.04266},
  year={2023},
  url = {https://doi.org/10.48550/arXiv.2311.04266}
}

@article{Preti2025,
  title={Single-fluid model for rotating annular supersolids and its experimental implications},
  author={Preti, Niccolò and Antolini, Nicolò and Drevon, Charles and Lombardi, Pietro and Fioretti, Andrea and Gabbanini, Carlo and Ferioli, Giovanni and Modugno, Giovanni and Biagioni, Giulio},
  journal={arXiv:2510.26753},
  year={2025},
  url = {https://arxiv.org/abs/2510.26753}
}

@article{Ding2025,
  title = {Interaction-induced reentrance of Bose glass and quench dynamics of Bose gases in twisted bilayer and quasicrystal optical lattices},
  author = {Ding, Shi-Hao and Lang, Li-Jun and Zhu, Qizhong and He, Liang},
  journal = {Phys. Rev. A},
  volume = {112},
  issue = {3},
  pages = {033322},
  numpages = {10},
  year = {2025},
  month = {Sep},
  publisher = {American Physical Society},
  doi = {10.1103/fvny-58kf},
  url = {https://link.aps.org/doi/10.1103/fvny-58kf}
}

@article{Trypogeorgos2025,
  title        = {Emerging supersolidity in photonic-crystal polariton condensates},
  author       = {Trypogeorgos, Dimitrios and Gianfrate, Antonio and Landini, Manuele and Nigro, Davide and Gerace, Dario and Carusotto, Iacopo and Riminucci, Fabrizio and Baldwin, Kirk W. and Pfeiffer, Loren N. and Martone, Giovanni I. and De Giorgi, Milena and Ballarini, Dario and Sanvitto, Daniele},
  journal      = {Nature},
  volume       = {639},
  pages        = {337--341},
  year         = {2025},
  publisher    = {Nature Publishing Group},
  doi          = {10.1038/s41586-025-08494-2},
}

@article{Manni2011,
  title = {Spontaneous Pattern Formation in a Polariton Condensate},
  author = {Manni, F. and Lagoudakis, K. G. and Liew, T. C. H. and Andr\'e, R. and Deveaud-Pl\'edran, B.},
  journal = {Phys. Rev. Lett.},
  volume = {107},
  issue = {10},
  pages = {106401},
  numpages = {4},
  year = {2011},
  month = {Sep},
  publisher = {American Physical Society},
  doi = {10.1103/PhysRevLett.107.106401},
  url = {https://link.aps.org/doi/10.1103/PhysRevLett.107.106401}
}

@article{sindik2024,
  title = {Sound, Superfluidity, and Layer Compressibility in a Ring Dipolar Supersolid},
  author = {\ifmmode \check{S}\else \v{S}\fi{}indik, Marija and Zawi\ifmmode \acute{s}\else \'{s}\fi{}lak, Tomasz and Recati, Alessio and Stringari, Sandro},
  journal = {Phys. Rev. Lett.},
  volume = {132},
  issue = {14},
  pages = {146001},
  numpages = {7},
  year = {2024},
  month = {Apr},
  publisher = {American Physical Society},
  doi = {10.1103/PhysRevLett.132.146001},
  url = {https://link.aps.org/doi/10.1103/PhysRevLett.132.146001}
}

@article{Abad2012a,
  title = {Dipolar condensates confined in a toroidal trap: Ground state and vortices},
  author = {Abad, M. and Guilleumas, M. and Mayol, R. and Pi, M. and Jezek, D. M.},
  journal = {Phys. Rev. A},
  volume = {81},
  issue = {4},
  pages = {043619},
  numpages = {8},
  year = {2010},
  month = {Apr},
  publisher = {American Physical Society},
  doi = {10.1103/PhysRevA.81.043619},
  url = {https://link.aps.org/doi/10.1103/PhysRevA.81.043619}
}

@article{Abad2012b,
  title = {Phase slippage and self-trapping in a self-induced bosonic {Josephson} junction},
  author = {Abad, M. and Guilleumas, M. and Mayol, R. and Pi, M. and Jezek, D. M.},
  journal = {Phys. Rev. A},
  volume = {84},
  issue = {3},
  pages = {035601},
  numpages = {4},
  year = {2011},
  month = {Sep},
  publisher = {American Physical Society},
  doi = {10.1103/PhysRevA.84.035601},
  url = {https://link.aps.org/doi/10.1103/PhysRevA.84.035601}
}

@article{ HiggsMukherjee2025,
  title = {Quantum Carpets of {Higgs} Quasiparticles in a Supersolid},
  author = {Mukherjee, K. and Schubert, M. and Klemt, R. and Bland, T. and Pfau, T. and Reimann, S. M.},
  journal = {Phys. Rev. Lett.},
  volume = {135},
  issue = {22},
  pages = {223402},
  numpages = {8},
  year = {2025},
  month = {Nov},
  publisher = {American Physical Society},
  doi = {10.1103/6d1g-671p},
  url = {https://link.aps.org/doi/10.1103/6d1g-671p}
}

@article{Platt2024,
  title={Sound waves and fluctuations in one-dimensional supersolids},
  author={Platt, LM and Baillie, D and Blakie, PB},
  journal={Physical Review A},
  volume={110},
  number={2},
  pages={023320},
  year={2024},
  publisher={APS},
url = {https://doi.org/10.1103/PhysRevA.110.023320}
}

@article{Klaus2022,
author={Klaus, Lauritz
and Bland, Thomas
and Poli, Elena
and Politi, Claudia
and Lamporesi, Giacomo
and Casotti, Eva
and Bisset, Russell N.
and Mark, Manfred J.
and Ferlaino, Francesca},
title={Observation of vortices and vortex stripes in a dipolar condensate},
journal={Nature Physics},
year={2022},
month={Dec},
day={01},
volume={18},
number={12},
pages={1453-1458},
issn={1745-2481},
doi={10.1038/s41567-022-01793-8},
url={https://doi.org/10.1038/s41567-022-01793-8}
}

@article{Gross1957,
  title = {Unified Theory of Interacting Bosons},
  author = {Gross, Eugene P.},
  journal = {Phys. Rev.},
  volume = {106},
  issue = {1},
  pages = {161--162},
  numpages = {0},
  year = {1957},
  month = {Apr},
  publisher = {American Physical Society},
  doi = {10.1103/PhysRev.106.161},
  url = {https://link.aps.org/doi/10.1103/PhysRev.106.161}
}

@article{Yang1962,
  title = {Concept of Off-Diagonal Long-Range Order and the Quantum Phases of Liquid He and of Superconductors},
  author = {Yang, C. N.},
  journal = {Rev. Mod. Phys.},
  volume = {34},
  issue = {4},
  pages = {694--704},
  numpages = {0},
  year = {1962},
  month = {Oct},
  publisher = {American Physical Society},
  doi = {10.1103/RevModPhys.34.694},
  url = {https://link.aps.org/doi/10.1103/RevModPhys.34.694}
}

@article{Boninsegni2012,
  title = {{Colloquium: Supersolids: What and where are they?}},
  author = {Boninsegni, Massimo and Prokof'ev, Nikolay V.},
  journal = {Rev. Mod. Phys.},
  volume = {84},
  issue = {2},
  pages = {759--776},
  numpages = {0},
  year = {2012},
  month = {May},
  publisher = {American Physical Society},
  doi = {10.1103/RevModPhys.84.759},
  url = {https://link.aps.org/doi/10.1103/RevModPhys.84.759}
}

@article{Tengstrand2023,
  title = {{Toroidal dipolar supersolid with a rotating weak link}},
  author = {{M. N. Tengstrand, and P. St\"urmer, and J. Ribbing, and S. M. Reimann}},
  journal = {Phys. Rev. A},
  volume = {107},
  issue = {6},
  pages = {063316},
  numpages = {7},
  year = {2023},
  month = {Jun},
  publisher = {American Physical Society},
  doi = {10.1103/PhysRevA.107.063316},
  url = {https://link.aps.org/doi/10.1103/PhysRevA.107.063316}
}

@article{Schmidt2021OblateRoton,
  title = {{Roton Excitations in an Oblate Dipolar Quantum Gas}},
  author = {{Schmidt, J.-N. and Hertkorn, J. and Guo, M. and B\"ottcher, F. and Schmidt, M. and Ng, K. S. H. and Graham, S. D. and Langen, T. and Zwierlein, M. and Pfau, T.}},
  journal = {Phys. Rev. Lett.},
  volume = {126},
  issue = {19},
  pages = {193002},
  numpages = {6},
  year = {2021},
  month = {May},
  publisher = {American Physical Society},
  doi = {10.1103/PhysRevLett.126.193002},
  url = {https://link.aps.org/doi/10.1103/PhysRevLett.126.193002}
}

@article{Natale2019,
  title = {Excitation Spectrum of a Trapped Dipolar Supersolid and Its Experimental Evidence},
  author = {Natale, G. and van Bijnen, R. M. W. and Patscheider, A. and Petter, D. and Mark, M. J. and Chomaz, L. and Ferlaino, F.},
  journal = {Phys. Rev. Lett.},
  volume = {123},
  issue = {5},
  pages = {050402},
  numpages = {6},
  year = {2019},
  month = {Aug},
  publisher = {American Physical Society},
  doi = {10.1103/PhysRevLett.123.050402},
  url = {https://link.aps.org/doi/10.1103/PhysRevLett.123.050402}
}

@article{Hertkorn2019,
  title = {Fate of the Amplitude Mode in a Trapped Dipolar Supersolid},
  author = {Hertkorn, J. and B\"ottcher, F. and Guo, M. and Schmidt, J. N. and Langen, T. and B\"uchler, H. P. and Pfau, T.},
  journal = {Phys. Rev. Lett.},
  volume = {123},
  issue = {19},
  pages = {193002},
  numpages = {6},
  year = {2019},
  month = {Nov},
  publisher = {American Physical Society},
  doi = {10.1103/PhysRevLett.123.193002},
  url = {https://link.aps.org/doi/10.1103/PhysRevLett.123.193002}
}

@article{Sanchez-Baena2023,
author={S{\'a}nchez-Baena, J.
and Politi, C.
and Maucher, F.
and Ferlaino, F.
and Pohl, T.},
title={Heating a dipolar quantum fluid into a solid},
journal={Nature Communications},
year={2023},
month={Apr},
day={04},
volume={14},
number={1},
pages={1868},
issn={2041-1723},
doi={10.1038/s41467-023-37207-3},
url={https://doi.org/10.1038/s41467-023-37207-3}
}

@article{Ilg2023,
  title = {Ground-state stability and excitation spectrum of a one-dimensional dipolar supersolid},
  author = {Ilg, Tobias and B\"uchler, Hans Peter},
  journal = {Phys. Rev. A},
  volume = {107},
  issue = {1},
  pages = {013314},
  numpages = {12},
  year = {2023},
  month = {Jan},
  publisher = {American Physical Society},
  doi = {10.1103/PhysRevA.107.013314},
  url = {https://link.aps.org/doi/10.1103/PhysRevA.107.013314}
}

@article{Hertkorn2021Spectrum2D,
  title = {Supersolidity in Two-Dimensional Trapped Dipolar Droplet Arrays},
  author = {Hertkorn, J. and Schmidt, J.-N. and Guo, M. and B\"ottcher, F. and Ng, K. S. H. and Graham, S. D. and Uerlings, P. and B\"uchler, H. P. and Langen, T. and Zwierlein, M. and Pfau, T.},
  journal = {Phys. Rev. Lett.},
  volume = {127},
  issue = {15},
  pages = {155301},
  numpages = {6},
  year = {2021},
  month = {Oct},
  publisher = {American Physical Society},
  doi = {10.1103/PhysRevLett.127.155301},
  url = {https://link.aps.org/doi/10.1103/PhysRevLett.127.155301}
}

@article{Chomaz2023,
doi = {10.1088/1361-6633/aca814},
url = {https://dx.doi.org/10.1088/1361-6633/aca814},
year = {2023},
month = {jan},
publisher = {IOP Publishing},
volume = {86},
number = {2},
pages = {026401},
author = {Lauriane Chomaz and Igor Ferrier-Barbut and Francesca Ferlaino and Bruno Laburthe-Tolra and Benjamin L Lev and Tilman Pfau},
title = {Dipolar physics: a review of experiments with magnetic quantum gases},
journal = {Reports on Progress in Physics}
}

@article{mukherjee2023droplets,
  title={Droplets and supersolids in ultra-cold atomic quantum gases},
  author={Mukherjee, K and Cardinale, T Arnone and Chergui, L and St{\"u}rmer, P and Reimann, SM},
  journal={The European Physical Journal Special Topics},
  volume={232},
  number={20},
  pages={3417--3433},
  year={2023},
  publisher={Springer},
  url = {https://doi.org/10.1140/epjs/s11734-023-00991-6}
}

@article{Del_pace2022,
  title = {Imprinting Persistent Currents in Tunable Fermionic Rings},
  author = {Del Pace, G. and Xhani, K. and Muzi Falconi, A. and Fedrizzi, M. and Grani, N. and Hernandez Rajkov, D. and Inguscio, M. and Scazza, F. and Kwon, W. J. and Roati, G.},
  journal = {Phys. Rev. X},
  volume = {12},
  issue = {4},
  pages = {041037},
  numpages = {16},
  year = {2022},
  month = {Dec},
  publisher = {American Physical Society},
  doi = {10.1103/PhysRevX.12.041037},
  url = {https://link.aps.org/doi/10.1103/PhysRevX.12.041037}
}

@article{Bisset2016,
  title = {Ground-state phase diagram of a dipolar condensate with quantum fluctuations},
  author = {Bisset, R. N. and Wilson, R. M. and Baillie, D. and Blakie, P. B.},
  journal = {Phys. Rev. A},
  volume = {94},
  issue = {3},
  pages = {033619},
  numpages = {10},
  year = {2016},
  month = {Sep},
  publisher = {American Physical Society},
  doi = {10.1103/PhysRevA.94.033619},
  url = {https://link.aps.org/doi/10.1103/PhysRevA.94.033619}
}

@article{Lima2012,
	author = {Lima, A. R. P. and Pelster, A.},
	doi = {10.1103/PhysRevA.86.063609},
	isbn = {9780124078666},
	issn = {10502947},
	journal = {Phys. Rev. A},
	number = {6},
	pages = {063609},
	title = {{Beyond mean-field low-lying excitations of dipolar Bose gases}},
	volume = {86},
	year = {2012}
}

@article{Lima2011,
  title = {Quantum fluctuations in dipolar {Bose} gases},
  author = {Lima, Aristeu R. P. and Pelster, Axel},
  journal = {Phys. Rev. A},
  volume = {84},
  issue = {4},
  pages = {041604},
  numpages = {4},
  year = {2011},
  month = {Oct},
  publisher = {American Physical Society},
  doi = {10.1103/PhysRevA.84.041604},
  url = {https://link.aps.org/doi/10.1103/PhysRevA.84.041604}
}

@article{Mukherjee_BHS_2022,
  author  = {Mukherjee, Biswaroop and Shaffer, Airlia and Patel, Parth B. and Yan, Zhenjie and Wilson, Cedric C. and Crépel, Valentin and Fletcher, Richard J. and Zwierlein, Martin},
  title   = {Crystallization of Bosonic Quantum {Hall} States in a Rotating Quantum Gas},
  journal = {Nature},
  volume  = {601},
  pages   = {58--62},
  year    = {2022},
  doi     = {10.1038/s41586-021-04202-9},
  url     = {https://www.nature.com/articles/s41586-021-04202-9}
}

@article{Yao2024,
  author  = {Yao, Ruixiao and Chi, Sungjae and Mukherjee, Biswaroop and Shaffer, Airlia and Zwierlein, Martin and Fletcher, Richard J.},
  title   = {Observation of Chiral Edge Transport in a Rapidly Rotating Quantum Gas},
  journal = {Nat. Phys.},
  volume  = {20},
  pages   = {1726--1731},
  year    = {2024},
  doi     = {10.1038/s41567-024-02617-7},
  url     = {https://doi.org/10.1038/s41567-024-02617-7}
}

@article{cai2022_Fermi,
  title = {Persistent Currents in Rings of Ultracold Fermionic Atoms},
  author = {Cai, Yanping and Allman, Daniel G. and Sabharwal, Parth and Wright, Kevin C.},
  journal = {Phys. Rev. Lett.},
  volume = {128},
  issue = {15},
  pages = {150401},
  numpages = {7},
  year = {2022},
  month = {Apr},
  publisher = {American Physical Society},
  doi = {10.1103/PhysRevLett.128.150401},
  url = {https://link.aps.org/doi/10.1103/PhysRevLett.128.150401}
}

@article{Guo2019,
author = {Guo, Mingyang and B{\"{o}}ttcher, Fabian and Hertkorn, Jens and Schmidt, Jan-Niklas and Wenzel, Matthias and B{\"{u}}chler, Hans Peter and Langen, Tim and Pfau, Tilman},
doi = {10.1038/s41586-019-1569-5},
issn = {1476-4687},
journal = {Nature},
number = {7778},
pages = {386--389},
title = {{The low-energy Goldstone mode in a trapped dipolar supersolid}},
url = {https://doi.org/10.1038/s41586-019-1569-5},
volume = {574},
year = {2019}
}

@article{Tengstrand2021,
	title = {Persistent currents in toroidal dipolar supersolids},
	author = {Tengstrand, M. Nilsson and Boholm, D. and Sachdeva, R. and Bengtsson, J. and Reimann, S. M.},
	journal = {Phys. Rev. A},
	volume = {103},
	issue = {1},
	pages = {013313},
	numpages = {7},
	year = {2021},
	month = {Jan},
	publisher = {American Physical Society},
	doi = {10.1103/PhysRevA.103.013313},
	url = {https://link.aps.org/doi/10.1103/PhysRevA.103.013313}
}

@article{Andreev1969,
	title={{Quantum Theory of Defects In Cystals}}, 
	author={Andreev, A. F. and Lifshitz, I. M.},
	journal = {J. Exp. Theo. Phys.},
	volume = {56},
	issue = {6},
	pages = {2057--2068},
	year = {1969},
	month = {Jun},
	publisher = {Soviet Physics},
	url = {http://www.jetp.ac.ru/cgi-bin/e/index/e/29/6/p1107?a=list}
}

@article{Norcia2021,
  title={Two-dimensional supersolidity in a dipolar quantum gas},
  author={Norcia, Matthew A and Politi, Claudia and Klaus, Lauritz and Poli, Elena and Sohmen, Maximilian and Mark, Manfred J and Bisset, Russell N and Santos, Luis and Ferlaino, Francesca},
  journal={Nature},
  volume={596},
  number={7872},
  pages={357--361},
  year={2021},
  publisher={Nature Publishing Group UK London},
url = {https://doi.org/10.1038/s41586-021-03725-7}
}

@article{Schmitt2016,
	author={Schmitt, M.
	and Wenzel, M.
	and B{\"o}ttcher, F.
	and Ferrier-Barbut, I.
	and Pfau, T.},
	title={{Self-bound droplets of a dilute magnetic quantum liquid}},
	journal={Nature},
	year={2016},
	month={Nov},
	day={01},
	volume={539},
	number={7628},
	pages={259-262},
	issn={1476-4687},
	doi={10.1038/nature20126},
	url={https://doi.org/10.1038/nature20126}
}

@article{Chomaz2016,
	title = {{Quantum-Fluctuation-Driven Crossover from a Dilute Bose-Einstein Condensate to a Macrodroplet in a Dipolar Quantum Fluid}},
	author = {Chomaz, L. and Baier, S. and Petter, D. and Mark, M. J. and W\"achtler, F. and Santos, L. and Ferlaino, F.},
	journal = {Phys. Rev. X},
	volume = {6},
	issue = {4},
	pages = {041039},
	numpages = {10},
	year = {2016},
	month = {Nov},
	publisher = {American Physical Society},
	doi = {10.1103/PhysRevX.6.041039},
	url = {https://link.aps.org/doi/10.1103/PhysRevX.6.041039}
}

@article{Wachtler2016,
	title = {{Ground-state properties and elementary excitations of quantum droplets in dipolar Bose-Einstein condensates}},
	author = {W\"achtler, F. and Santos, L.},
	journal = {Phys. Rev. A},
	volume = {94},
	issue = {4},
	pages = {043618},
	numpages = {7},
	year = {2016},
	month = {Oct},
	publisher = {American Physical Society},
	doi = {10.1103/PhysRevA.94.043618},
	url = {https://link.aps.org/doi/10.1103/PhysRevA.94.043618}
}

@article{Roccuzzo2020,
	title = {{Rotating a Supersolid Dipolar Gas}},
	author = {Roccuzzo, S. M. and Gallem\'{\i}, A. and Recati, A. and Stringari, S.},
	journal = {Phys. Rev. Lett.},
	volume = {124},
	issue = {4},
	pages = {045702},
	numpages = {6},
	year = {2020},
	month = {Jan},
	publisher = {American Physical Society},
	doi = {10.1103/PhysRevLett.124.045702},
	url = {https://link.aps.org/doi/10.1103/PhysRevLett.124.045702}
}

@article{Chester1970,
	title = {{Speculations on Bose-Einstein Condensation and Quantum Crystals}},
	author = {Chester, G. V.},
	journal = {Phys. Rev. A},
	volume = {2},
	issue = {1},
	pages = {256--258},
	numpages = {0},
	year = {1970},
	month = {Jul},
	publisher = {American Physical Society},
	doi = {10.1103/PhysRevA.2.256},
	url = {https://link.aps.org/doi/10.1103/PhysRevA.2.256}
}

@article{Li2017,
  title = {A stripe phase with supersolid properties in spin-orbit-coupled
   Bose-Einstein condensates},
  author = {Li, Jun-Ru and Lee, Jeongwon and Huang, Wujie and Burchesky, Sean and
   Shteynas, Boris and Top, Furkan Cagri and Jamison, Alan O. and Ketterle,
   Wolfgang},
  journal = {Nature},
  volume = {543},
  pages = {91–94},
  year = {2017},
  month = {03},
  doi = {10.1038/nature21431},
  url = {https://www.nature.com/articles/nature21431}
}

@article{Leonard2017,
  title = {A stripe phase with supersolid properties in spin-orbit-coupled
   Bose-Einstein condensates},
  author = {Leonard, Julian and Morales, Andrea and Zupancic, Philip and Esslinger,
   Tilman and Donner, Tobias},
  journal = {Nature},
  volume = {543},
  pages = {87-90},
  year = {2017},
  month = {03},
  doi = {10.1038/nature21067},
  url = {https://www.nature.com/articles/nature21067}
}

@article{Chomaz2019,
	title = {Long-Lived and Transient Supersolid Behaviors in Dipolar Quantum Gases},
	author = {Chomaz, L. and Petter, D. and Ilzh\"ofer, P. and Natale, G. and Trautmann, A. and Politi, C. and Durastante, G. and van Bijnen, R. M. W. and Patscheider, A. and Sohmen, M. and Mark, M. J. and Ferlaino, F.},
	journal = {Phys. Rev. X},
	volume = {9},
	issue = {2},
	pages = {021012},
	numpages = {12},
	year = {2019},
	month = {Apr},
	publisher = {American Physical Society},
	doi = {10.1103/PhysRevX.9.021012},
	url = {https://link.aps.org/doi/10.1103/PhysRevX.9.021012}
}

@article{Casotti2024,
  title = {Observation of vortices in a dipolar supersolid},
  author = {Casotti, Eva and Poli, Elena and Klaus, Lauritz and Litvinov, Andrea and Ulm, Clemens and Politi, Claudia and Mark, Manfred J. and Bland, Thomas and Ferlaino, Francesca},
  journal = {Nature},
  volume = {635},
  number = {8038},
  pages = {327--331},
  year = {2024},
  publisher = {Nature Publishing Group},
  doi = {10.1038/s41586-024-08149-7}
}

@article{Pomeau1994,
  title = {Dynamics of a model of supersolid},
  author = {Pomeau, Yves and Rica, Sergio},
  journal = {Phys. Rev. Lett.},
  volume = {72},
  issue = {15},
  pages = {2426--2429},
  numpages = {0},
  year = {1994},
  month = {Apr},
  publisher = {American Physical Society},
  doi = {10.1103/PhysRevLett.72.2426},
  url = {https://link.aps.org/doi/10.1103/PhysRevLett.72.2426}
}

@article{Poli2024Excitations2D,
  title = {Excitations of a two-dimensional supersolid},
  author = {Poli, E. and Baillie, D. and Ferlaino, F. and Blakie, P. B.},
  journal = {Phys. Rev. A},
  volume = {110},
  issue = {5},
  pages = {053301},
  numpages = {10},
  year = {2024},
  month = {Nov},
  publisher = {American Physical Society},
  doi = {10.1103/PhysRevA.110.053301},
  url = {https://link.aps.org/doi/10.1103/PhysRevA.110.053301}
}

@article{kirkby2024,
  title = {Excitations of a Binary Dipolar Supersolid},
  author = {Kirkby, W. and Lee, Au-Chen and Baillie, D. and Bland, T. and Ferlaino, F. and Blakie, P. B. and Bisset, R. N.},
  journal = {Phys. Rev. Lett.},
  volume = {133},
  issue = {10},
  pages = {103401},
  numpages = {7},
  year = {2024},
  month = {Sep},
  publisher = {American Physical Society},
  doi = {10.1103/PhysRevLett.133.103401},
  url = {https://link.aps.org/doi/10.1103/PhysRevLett.133.103401}
}

@article{Schmidt2021,
  title = {Roton Excitations in an Oblate Dipolar Quantum Gas},
  author = {Schmidt, J.-N. and Hertkorn, J. and Guo, M. and B\"ottcher, F. and Schmidt, M. and Ng, K. S. H. and Graham, S. D. and Langen, T. and Zwierlein, M. and Pfau, T.},
  journal = {Phys. Rev. Lett.},
  volume = {126},
  issue = {19},
  pages = {193002},
  numpages = {6},
  year = {2021},
  month = {May},
  publisher = {American Physical Society},
  doi = {10.1103/PhysRevLett.126.193002},
  url = {https://link.aps.org/doi/10.1103/PhysRevLett.126.193002}
}

@article{Pitaevskii1984,
  title = {Layered structure of {Bose} crystals},
  author = {Pitaevskii, L. P.},
  journal = {JETP Lett.},
  volume = {39},
  number = {9},
  pages = {511--514},
  year = {1984},
  note = {[Pis'ma Zh. Eksp. Teor. Fiz. 39, 423 (1984)]}
}

\section{End Matter}
\prlsection{Appendix A} In this section, we describe the numerical methods employed to obtain the ground states and dynamical properties presented in the main text. All simulations are performed at zero temperature within the mean-field framework of the eGPE~\cite{Wachtler2016,Ferrier-Barbut2016,Chomaz2016,Bisset2016}. The time evolution of the condensate wavefunction $\Psi(\mathbf{r},t)$ in the rotating frame is governed by:
\begin{equation}
i\hbar \frac{\partial \Psi(\mathbf{r},t)}{\partial t} = \left( \hat{\mathcal{H}}- \Omega \hat{L}_z\right) \Psi(\mathbf{r},t),
\end{equation}
where the eGPE operator $\hat{\mathcal{H}}$ is defined as:
\begin{align}
\label{eGPE_full}
\hat{\mathcal{H}} = {} &
-\frac{\hbar^2}{2M}\nabla^2 + V_{\rm trap}(\mathbf{r}) + g |\Psi|^2 + g_{\rm QF} |\Psi|^3 \nonumber \\
& + \int d^3\mathbf{r}'\, U_{\mathrm{dd}}(\mathbf{r}-\mathbf{r}') |\Psi(\mathbf{r}',t)|^2.
\end{align}
The external potential $V_{\rm trap}(\mathbf{r})$ accounts for the static confinement, which in this work is either a toroidal or an oblate harmonic potential. Quantum fluctuations are incorporated via the LHY correction with the coefficient $g_{\rm QF} = \frac{32}{3} g \sqrt{a_s^3/\pi} (1 + \frac{3}{2}\epsilon_{\mathrm{dd}}^2)$, where $\epsilon_{\mathrm{dd}} = a_{\mathrm{dd}}/a_s$.

The corresponding energy functional associated with the eGPE, the minimization of which yields the stationary states, is given by $E^{\prime}[\Psi]=E[\Psi]-\Omega L_z$ with
\begin{equation}
\begin{split}
E[\Psi] = & \int d^3\mathbf{r} \bigg[ \frac{\hbar^2}{2M} |\nabla \Psi|^2 + V_{\rm trap}(\mathbf{r}) |\Psi|^2 + \frac{g}{2} |\Psi|^4 \\
& + \frac{2}{5} g_{\rm QF} |\Psi|^5 \bigg] + \frac{1}{2} \iint d^3\mathbf{r} \, d^3\mathbf{r}' \\
& \times |\Psi(\mathbf{r})|^2 U_{\rm dd}(\mathbf{r}-\mathbf{r}') |\Psi(\mathbf{r}')|^2.
\end{split}
\end{equation}

For numerical implementation, the eGPE is rescaled into a dimensionless form using a characteristic length scale $l_s=10^{-6} \, \rm m$ and time scale $t_s = M l_s^2 / \hbar$. Ground states are obtained via imaginary-time propagation using a split-step Fourier spectral method. To analyze the dynamical response and excitation spectra, we perform subsequent real-time propagation with a time step of $\Delta t = 10^{-4}\,t_s$. The nonlocal dipolar term is evaluated efficiently as a convolution in momentum space, utilizing a spherical cutoff for the interaction kernel to eliminate aliasing effects and spurious interactions between periodic images.

Simulations are conducted in a three-dimensional box with dimensions $L_x = L_y = L_z = 22\,\mu\mathrm{m}$. We confirmed numerical convergence by comparing results across grid resolutions of $128 \times 128 \times 64$ and $256 \times 256 \times 64$, noting that the accuracy of the BdG excitation frequencies depends primarily on the physical box size. 

To model the dynamics of the rotation-induced supersolid state, we include a phenomenological damping term in the eGPE:
\begin{equation}
(i - \gamma)\hbar \frac{\partial \Psi}{\partial t} = \left( \hat{\mathcal{H}}- \Omega \hat{L}_z\right) \,\Psi,
\end{equation}
where $\gamma \ll 1$ is a small dimensionless damping parameter. This term accounts for dissipative processes and facilitates the system's relaxation toward stable dynamical configurations. For the real-time dynamics discussed in the main text, we have used $\gamma = 0.035$ for the toroidal trap and $\gamma = 0.02$ for the oblate harmonic trap. At $t = 0$, we also introduce a small perturbation to account for experimental inhomogeneities and to break the angular symmetry.

\prlsection{Appendix B} To gain analytical insight into the excitation spectrum under rotation, we consider a superfluid confined to a narrow ring of radius $R$ satisfying $R \gg \sqrt{\hbar/M\omega_r}$. We derive the excitation energies as a function of the external rotation $\Omega$ and the ground-state winding number $q$ by assuming the solutions for the non-rotating case ($\Omega=q=0$) are known. This reference set of solutions is denoted by:
\begin{align}
    \left(\tilde{\omega}_m, \tilde{u}_m e^{im\phi}, \tilde{v}_m e^{im\phi}\right),
\end{align}
where $\tilde{u}_m$ and $\tilde{v}_m$ represent the quasiparticle amplitudes in the transversal $(r,z)$ directions. 

We now demonstrate that the vector
\begin{figure}[t]
	\centering
	\includegraphics[width = 0.5\textwidth]{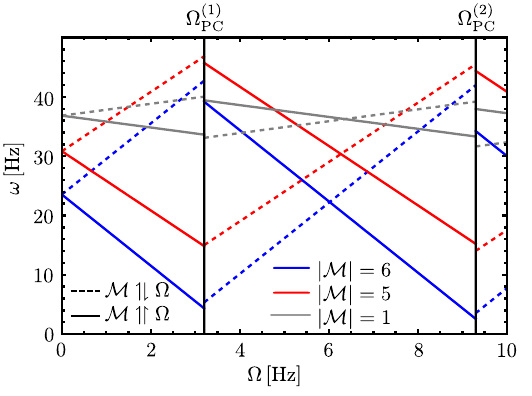}
    \vspace{-0.8cm}
\caption{\textbf{Mode-Swapping in a rotating toroidal Superfluid.} 
The energies of the six lowest excitations ($\mathcal{M}=\pm6,\pm 5,\pm 1$) for the system from Fig.2 in the main text with $a_s=93a_0$. The modes propagating in the direction of $\Omega$ are shown as solid lines, whereas we indicate modes propagating antiparallel to $\Omega$ as dashed lines. The vertical black lines indicate the critical rotation frequencies at which persistent currents are generated, corresponding to mode swapping.}
\label{Sup1}
\end{figure}
\begin{align}
\begin{pmatrix} u \\ v \end{pmatrix} = \begin{pmatrix} \tilde{u}_\mathcal{M} e^{i m \phi} \\ \tilde{v}_\mathcal{M} e^{i(m-2q)\phi} \end{pmatrix}
\end{align}
is a valid solution to the BdG equations and leads to the dispersion relation utilized in the main text. We begin by employing the tight-confinement approximation for the gradient operator, $\nabla_\phi \approx \frac{1}{R} \frac{\partial}{\partial \phi}$. Furthermore, we approximate the ground-state wavefunction as:
\begin{align}
    \Psi(\mathbf{r}) \approx |\tilde{\Psi}(r,z)| e^{iq\phi},
\end{align}
assuming that the radial density profile is approximately independent of the circulation $q$. Under these conditions, the chemical potential can be expressed as:
\begin{align}
    \mu = \tilde{\mu} + \frac{q^2 \hbar^2}{2MR^2} - \Omega \hbar q,
\end{align}
where $\tilde{\mu}$ is the chemical potential for the non-rotating system ($\Omega=q=0$). 

Inserting these expressions into the BdG equations and performing straightforward algebraic manipulation yields the following expression for the dispersion relation $\omega_{\mathcal{M}}^{(q)}$:
\begin{align}
    \omega_\mathcal{M}^{(q)}(\Omega) = \tilde{\omega}_\mathcal{M} + \frac{m q \hbar}{MR^2} - \frac{q^2 \hbar}{MR^2} + \Omega(q-m).
\end{align}
By introducing the fluctuation momentum $\mathcal{M} = m - q$, we recover the Galilean-shifted dispersion relation:
\begin{align}
    \omega_\mathcal{M}^{(q)}(\Omega) = \tilde{\omega}_\mathcal{M} - \mathcal{M} \left( \Omega - \frac{q \hbar}{MR^2} \right) = \tilde{\omega}_\mathcal{M} - \mathcal{M} \Omega_{\rm eff}.
\end{align}

It is important to note that the derivation above strictly holds when the radial and angular motions are decoupled ($R \gg \sqrt{\hbar/M\omega_r}$). In our numerical simulations, this condition is not perfectly satisfied, and the particle density $|\Psi|^2$ acquires a slight $q$-dependence due to radial centrifugal shifts, see Fig. \ref{Sup1}. Due to the radial coupling, the energies are not continuous, $\omega^{(q-1)}_{\mathcal{M}}(\Omega_{\rm PC}^{(q)})\neq \omega^{q}_{\mathcal{-M}}(\Omega_{\rm PC}^{(q)})$. However, we find that the general form of the dispersion remains robust if one replaces the intrinsic frequency with a $q$-dependent one, $\tilde{\omega}_\mathcal{M} \rightarrow \tilde{\omega}_\mathcal{M}^{(q)}$. This replacement allows us to construct the full spectrum by combining the numerically calculated excitation energies at different $q$ values with the analytical Doppler shift, rather than solving the full BdG problem for every value of $\Omega$.

This method remains valid as long as the radial trapping frequency dominates the rotation ($\omega_r \gg \Omega$). If this condition is violated, centrifugal effects would cause the density to depend explicitly on $\Omega$. Our numerical analysis confirms that such effects are negligible for the parameter regimes studied in the main text. Nevertheless, all results presented in the main text are cross-verified by solving the full, non-decoupled BdG problem to ensure that all rotational and interaction effects are rigorously captured.
\end{document}